\documentclass[aps,pra,twocolumn,showpacs,superscriptaddress,groupedaddress]{revtex4}  % for review and submission
\usepackage{graphicx}  % needed for figures
\usepackage{dcolumn}   % needed for some tables
\usepackage{bm}        % for math
\usepackage{amssymb}   % for math
\usepackage{amsmath}

\usepackage{hyperref}

\begin{document}

% The following information is for internal review, please remove them for submission
\widetext
%\leftline{Version xx as of \today}
%\leftline{Primary authors: Joe E. Physics}
%\leftline{To be submitted to (PRL, PRD-RC, PRD, PLB; choose one.)}
%\leftline{Comment to {\tt d0-run2eb-nnn@fnal.gov} by xxx, yyy}
%\centerline{\em D\O\ INTERNAL DOCUMENT -- NOT FOR PUBLIC DISTRIBUTION}

% the following line is for submission, including submission to the arXiv!!
%\hspace{5.2in} \mbox{Fermilab-Pub-04/xxx-E}

\title{Angular distributions in two-colour two-photon ionization of He}
\author{H.F. Rey}
\email{h.rey@qub.ac.uk}
%\author{J. S. Parker}
\author{H.W. van der Hart}
%\author{K. T. Taylor}

%\email{h.rey@qub.ac.uk}
%\homepage{http://stoa.usp.br/thschiavo}
\affiliation{Centre for Theoretical Atomic, Molecular and Optical Physics, School of Mathematics and Physics\\ Queen's University Belfast, Belfast BT7 1NN, United Kingdom}

%\input author_list.tex       % D0 authors (remove the first 3 lines
                             % of this file prior to submission, they
                             % contain a time stamp for the authorlist)
                             % (includes institutions and visitors)
\date{\today}

\begin{abstract}

We present R-Matrix with time dependence (RMT) calculations for the photoionization of helium
irradiated by an EUV laser pulse and an overlapping IR pulse with an emphasis on the anisotropy
parameters of the sidebands generated by the dressing laser field. We investigate how these
parameters depend on the amount of atomic structure included in the theoretical model for
two-photon ionization. To verify the accuracy of the RMT approach, our
theoretical results are compared with experiment. 

%An article usually includes an abstract, a concise summary of the work covered at length in the
%main body of the article. It is used for secondary publications and for information retrieval
%purposes.
%For PRL, the rule of thumb is that the abstract should be less than 8 lines and the text
%(excluding authors, abstract but including tables, figures and references) should be less
%than 4 pages (leave about 20 lines empty on page 4) in two-column format.
%PRL and PRD papers have to have PACS (Physics and Astronomy Classification Scheme) numbers.
%Please see {\tt http://www.aip.org/pacs/} for the numbers relevant to your paper. A set of
%standard references can be found at the end of this example paper.

\end{abstract}

\maketitle

\section{\label{sec:level1}  Introduction}
%\section{\label{sec:level1}First-level heading}
% sections are not used for PRL papers

Over the last 15 years significant progress has been made in the development of light sources,
capable of generating ultra-short pulses lasting just a fraction of a femtosecond  \cite{Iva09}.
One of the key experimental challenges has been the characterisation of these light pulses
\cite{Pau01}. By
overlapping the ultra-short pulse with an IR pulse during an ionization process, sidebands
are observed in photoelectron spectra. Examination of these sidebands allows the extraction
of the phase differences between the different harmonic constituents of the ultra-short
pulse \cite{Ven96}. Knowledge of these phase differences then allows the reconstruction
of the ultra-short light pulse.

Interest in sidebands created when photoionization by high-frequency light occurs in the
presence of an IR field is not just limited to the
characterisation of ultra-short light pulses. With the advent of free-electron
lasers operating at high photon energies \cite{A13,B13,C13},
the interplay between EUV photons and IR photons in the photoionization process \cite{Mey13}
can give valuable information about the time delay between the EUV and the IR pulse.

The experimental interest in sidebands generated by the addition of an IR field to an EUV
field should be complemented by theoretical investigation. A good description of these
sidebands is given by the soft-photon approximation (SPA) \cite{Maq07}, in which the angular
distribution of the ejected electron is modified by Bessel functions, depending on the number
of IR photons absorbed and the angle at which the electron is emitted with respect to the
polarization direction. This approximation applies when the energy of the IR photon is much
larger than the energy of electrons emitted by the EUV field by itself. When the energies of
the emitted electron and the IR photon are comparable, the ionic potential will affect the
motion of the outgoing electron, and the predictions of the SPA may not be as accurate.

In the present study, we investigate the effect of the residual ionic potential on the angular
distributions of sidebands generated during EUV+IR photoionization of He using the R-matrix
including time-dependence (RMT) approach \cite{Nik08,Lys11,Moo11}. The He atom has been chosen,
since the angular distributions of the photoelectron ejected in two-colour fields have been the
subject of experimental investigation, either when the EUV pulse is
sufficiently energetic to eject an electron from He
\cite{Mey10,Hab11}, or when the pulse excites one of the 1s electrons to an excite np state
\cite{Mon13,OKe13}. In addition, for He, it is straightforward to change the amount of
atomic structure retained in the calculations. A similar study of photo-electron angular
distributions has recently been performed
for Ar using the time-dependent R-matrix (TDRM) approach \cite{Hut13}, including a comparison
with experiment \cite{Hab09}, the SPA \cite{Maq07} and model potential anisotropy parameters
\cite{Tom01}. The present calculation thus also allows us to assess the computational
efficiency of the two different time-dependent R-matrix approaches.

The RMT approach was developed only a few years ago\cite{Nik08,Lys11,Moo11}, and has since
been applied to study time delays in photoionization of Ne \cite{Moo11b}, to two-photon
double ionization of He \cite{Har14}, and to IR-assisted photoionization of Ne$^+$ \cite{Har14b}.
Similar to the
TDRM approach \cite{Lys09}, it employs the standard R-matrix technique of separating configuration
space into an inner region and an outer region. However, the wavefunction is propagated using an
Arnoldi approach \cite{Par86,Smy98} in the RMT approach, compared to a Crank-Nicholson scheme in
the TDRM approach. Spatially, the TDRM approach applies a sequential R-matrix propagation scheme,
which restricts the degree of parallelisation to about 100-200 processors. This sequential step
does not occur within the RMT approach, and it can therefore be efficiently parallelised over
substantially larger numbers of processors \cite{Har14}.

In the present report, we will first describe the RMT approach and indicate the differences
between the RMT approach and the TDRM approach. We then apply the RMT approach to two-colour
two-photon ionization of He, and present the associated anisotropy parameters. We briefly
investigate how these parameters depend on the theoretical description of He. We compare
the obtained parameters to those obtained experimentally as well as the predictions from
the SPA. 

\section{The R-matrix including time approach}

The R-matrix including time (RMT) approach is a new  \textit{ab initio} method to solve
accurately the time dependent Schr\"{o}dinger equation (TDSE) for multi-electron atoms
in intense laser light \cite{Nik08,Lys11,Moo11}. The approach adopts the standard R-matrix
approach \cite{Bur11} of splitting configuration space into two distinct regions: an inner
region and an outer region. In the inner region, all $N+1$ electrons in the system
are contained within a sphere of radius $b$ around the nucleus of the atom. $N$ therefore
indicates the number of electrons left on the ion after a single electron has been ejected.
In the outer region $N$ electrons still remain within the sphere of radius $b$, but one of
the electrons has now left the sphere, so that its radial distance from the nucleus, $r_{N+1}$,
is larger than $b$.

In the inner region, a standard R-matrix basis expansion is used to describe the wave
function \cite{Bur11}. In this expansion, states of the system under investigation are
described in terms of antisymmetrised direct products of residual-ion states with a full
continuum basis for the outgoing electron. Additional correlation orbitals can be included
to improve the description of the system. All interactions are accounted for in the Hamiltonian,
including electron exchange and correlation effects between all pairs of electrons.
%The solution of the TDSE in the internal region uses a combination of standard R-matrix techniques
%and the Arnoldi propagator for the HELIUM %method~\cite{12} to obtain a solution to the
%TDSE and to propagate this solution forwards through a step  $\delta t$ in time. %\\

In the outer region, the wavefunction is described in terms of a direct product of residual-ion
states, coupled with the spin and angular momentum of the outgoing electron, with the radial
wavefunction for the outgoing electron. Since the outer electron is separated from the other
electrons, its wavefunction can be treated separately from the others, and hence exchange
effects can be neglected. In the RMT method, the radial wavefunction for the outer electron
is described in terms of a finite difference grid, similar to the approach pursued in the
HELIUM codes \cite{Smy98}. This is the first difference compared to
the TDRM approach \cite{Lys09}, in which the wave
function was described in terms of a very dense set of B-spline functions.

In sections \ref{sec:outer} and \ref{sec:inner}, we give a brief overview of the theory
underpinning the RMT approach. A full description of the approach is given in \cite{Moo11,Lys11}. 
To obtain the properties of the system under investigation, we start with the
time-dependent Schr\"odinger equation (TDSE),
\begin{eqnarray}
 i\frac{\partial }{\partial t}\Psi(\mathbf{X}_{N+1},t)=H_{N+1}(t)\Psi(\mathbf{X}_{N+1},t),
 \label{eq:tdse}
\end{eqnarray}
where $H_{N+1}$ is the full Hamiltonian for the system, given by:
\begin{equation}
H_{N+1}=\sum_{i=1}^{N+1}\left ( -\frac{1}{2}\bigtriangledown_{i}^{2}-\frac{Z}{r_{i}}+
\sum_{i>j=1}^{N+1}\frac{1}{r_{ij}}+\mathbf{E}(t)\cdot \sum_{i=1}^{N+1}\mathbf{r_{i}}\right ).
\label{eq:ham}
\end{equation}
In this equation $\mathbf{X}_{N+1}\equiv \boldsymbol{x}_{1},\boldsymbol{x}_{2}, \dots,
\boldsymbol{x}_{N+1}$ and $\boldsymbol{x}_{i}\equiv \boldsymbol{r}_{i}\sigma_{i}$, with
$\boldsymbol{r}_{i}$ and $\boldsymbol{\sigma}_{i}$ the position and spin vectors of the
$i^{\rm th}$ electron,
respectively. $Z$ indicates the nuclear charge and $\boldsymbol{E}(t)$ is the electric
field of the light pulse. Furthermore, $r_{ij}=\left |\boldsymbol{r}_{i}-\boldsymbol{r}_{j}\right|$.
 The nucleus has been taken at the origin of the coordinate system.

\subsection{The outer region}
\label{sec:outer}

In the outer region the $(N+1)$-electron wavefunction is expanded as follows \cite{Bur11}:
\begin{eqnarray}
\Psi(\mathbf{X}{_{N+1},t})=\sum_{p}\overline{\Phi}_p(\mathbf{X}_{N};\boldsymbol{\hat{r}},
\sigma_{N+1}) \frac{1}{r}F_{p}(r,t),
\end{eqnarray}
 where \textit{$r\equiv r_{N+1}$} is the radial distance of the $(N+1)^{\rm th}$ electron.
The channel functions $\overline{\Phi}_p$ are formed by coupling relevant states of the
residual $N$-electron ionic system $\Phi_{T}(X_{N})$ with angular and spin components of
the ejected-electron wavefunction. The functions \textit{$F_{p}(r,t)$} describe the
radial wavefunction of the outer electron in the $p^{\rm th}$ channel.

By left-projecting the TDSE (\ref{eq:tdse}) onto the channel functions
$\overline{\Phi}_p$, and integrating over all spatial and spin coordinates except $r$,
equation (\ref{eq:tdse}) can be rewritten in terms of a set of
coupled partial differential equations
for  $F_{p}(r,t)$ \cite{Lys11,Moo11}:
\begin{eqnarray}
	i\frac{\partial }{\partial t}F_{p}(r,t)& = & h_{II_{p}}(r)F_{p}(r,t)
	+\sum _{p'}[W_{E_{pp'}}(r)+W_{D_{pp'}}(t) \nonumber \\
	& + & W_{P_{pp'}}(r,t)]F_{p'}(r,t), \label{equation2}
\end{eqnarray}
where $h_{II_{p}}(r)$ describes the energy of the residual-ion state, the kinetic energy,
the screened nuclear attraction and centrifugal repulsion for
the outer electron, 
\begin{equation}
        \label{equation4}
			h_{II_{p}}(r)=-\frac{1}{2}\frac{d^2 }{d r^2}+
			\frac{l_{p}(l_{p}+1)}{2r^{2}}-\frac{Z-N}{r}+E_{p}.
\end{equation}
The other terms in equation (\ref{equation2}) describe the coupling between the different
channels. $W_{E_{pp'}}(r)$ describes the coupling due to long-range repulsion terms arising from
the residual electrons, excluding the screening term,
\begin{equation}
        \label{equation5}
			W_{E_{pp'}}(r)=\left \langle r^{-1}
			\mathit{\overline{\Phi}_{p}} \left |\sum_{j=1}^{N}
			\frac{1}{|\mathbf{r}-\mathbf{r_{j}}|}-\frac{N}{r} \right |
			r^{-1}\mathit{\overline{\Phi}_{p'}} \right \rangle.	
\end{equation}
$W_{D_{pp'}}(t)$ describes the coupling due to interaction between the light field and
the residual-ion states,
\begin{equation}
        \label{equation6}
        		W_{D_{pp'}}(t)=\left \langle r^{-1}
			\mathit{\overline{\Phi}_{p}}\left |
			\mathbf{E}(t)\cdot \sum_{i=1}^{N}\boldsymbol{r_{i}} \right |
			r^{-1}\mathit{\overline{\Phi}_{p'}} \right \rangle.
\end{equation}
Finally, $W_{P_{pp'}}(t)$ describes the coupling due to interaction between the light field and
the outer electron,
\begin{equation}
        \label{equation7}
		W_{P_{pp'}}(r,t)=\left \langle r^{-1}\mathit{\overline{\Phi}_{p}}
		\left |
		\mathbf{E}(t)\cdot\boldsymbol{r} \right |r^{-1}
		\mathit{\overline{\Phi}_{p'}} \right \rangle.
\end{equation}

\begin{figure}[tbh]
	\centering
		\includegraphics[width=8cm]{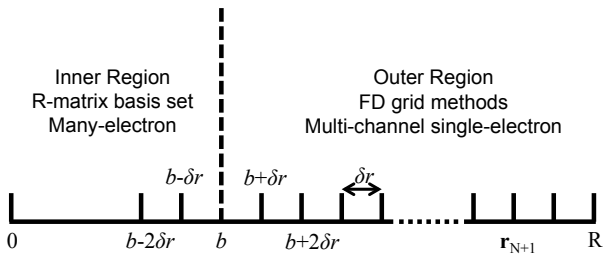}%[width=300pt]
	\caption{The R-matrix division-of-space concept. In the inner region a basis expansion
	of the wavefunction is chosen, while in the outer region a grid-based representation
	is considered. The vertical dashes indicate the grid points used in the calculation. The
	outer-region grid extrapolates into the inner region. At these points, the inner region
	wavefunction is projected onto the outer-region grid associated with each channel.
	The boundary of the inner region is at $r = b$ and the outer region boundary
	is at $r = R$.
	%The radial distance of the (\textit{N}+1)$^{\rm th}$
	%electron is denoted by $r$.
	}
  \label{fig:rmt_partition}
 \end{figure}

Equation (\ref{equation2}) describes a system of equations, which can be solved
efficiently using finite-difference techniques \cite{Lys11,Moo11}.
The radial
wavefunction for the outer electron, $F_{p}(r,t)$, is discretised on a radial grid as shown in figure
\ref{fig:rmt_partition}. The potential-energy terms and the angular-momentum repulsion term can be
determined directly at each grid point for each channel. The second derivative term in the
kinetic energy operator is implemented using a five-point
finite-difference rule. Near the boundary with the inner region, wavefunction information from
the inner region is required. We therefore extend the
finite difference grid into the inner region by $2N_A$ points, where $N_A$ is the order of
the Arnoldi propagator. Following the determination of the inner-region field-free eigenstates,
we project these inner-region eigenstates onto the channel functions.
Using the time-dependent coefficients $C_k(t)$ associated with the field-free inner-region
eigenstates
(see section \ref{sec:inner}), we can evaluate the time-dependent inner-region wavefunction on
the finite-difference grid extension into the inner region (eg. the points $b - \delta r$ and
$b - 2\delta r$ in figure \ref{fig:rmt_partition}). Once the wavefunction is known on these points,
all required orders of the kinetic energy operator can be evaluated accurately
in the outer region.
To propagate the outer region wavefunction in time from $t$ to $t + \delta t$ we employ the
Arnoldi propagator \cite{Smy98,Par86}.

The most demanding part in the solution of equation (\ref{equation2}) in the present approach
is the evaluation of the right-hand side. This is effectively a matrix-vector
multiplication, where the matrix will be sparse due to the angular-momentum constraints on
the individual matrix elements. On the other hand, solution of the outer-region equations in
the TDRM approach requires the (computationally demanding)
determination of time-dependent Green's functions across the entire outer region \cite{Lys09}.
In addition, the TDRM approach determines the updated wavefunction through an R-matrix
propagation scheme, which cannot be carried out in parallel. In the RMT approach, the
updated wavefunction is obtained through a purely local computational scheme, and can therefore
be parallelised over very large numbers of cores.  As a consequence, the RMT codes are
substantially faster than the TDRM codes, especially for large-scale problems.

\subsection{The inner region}
\label{sec:inner}

The time dependent $(N+1)$-electron wavefunction in the inner region $\Psi_I(X_{N+1},t)$ is
expanded over field-free eigenstates of the Hamiltonian in the inner
region $\psi_{k}(X_{N+1})$ as follows \cite{Bur11}:
\begin{equation}
        \label{equation14}
			\Psi_I(X_{N+1},t)=\sum_{k}C_{k}(t)\psi_k(X_{N+1}),
\end{equation}
where all $r_{i} \leq b$ and the $C_{k}(t)$ are the time-dependent expansion coefficients
associated with the field-free eigenstates.

The TDSE in the inner region is given by:
\begin{equation}
        \label{eq:tdinner}
			i\frac{\partial }{\partial t}\Psi(X_{N+1},t)=H_{N+1}(t)\Psi(X_{N+1},t),
\end{equation}
where the time-dependent Hamiltonian is the same as given in expression (\ref{eq:ham}).
However, in the inner region, the Hamiltonian $H_{N+1}(t)$ cannot be hermitian. Ionization
requires an electron to escape to infinity, and for this
to occur, some part of the wavefunction must have left the inner region. Hence the
total population in the inner region cannot be conserved. The non-hermiticity arises in
the evaluation of the kinetic energy operator at the boundary. In the
determination of the field-free inner-region eigenstates, however,
this non-hermiticity is compensated
for through the addition of a Bloch operator \cite{Bur11},
\begin{equation}
        \label{equation16}
		\mathcal{L}_{N+1}=\frac{1}{2}\sum_{i=1}^{N+1}\delta (r_{i}-b)
		\left ( \frac{d}{dr_{i}}-\frac{g_{0}-1}{r_{i}} \right ), 
\end{equation}
where the value of $g_{0}$ can, in principle, be chosen freely.

To propagate the inner-region wavefunction in time using the correct time-dependent Hamiltonian, we
therefore have to remove the Bloch operator \cite{Lys11,Moo11}:
\begin{equation}
        \label{equation17}
		i\frac{\partial }{\partial t}\Psi_{I}(X_{N+1},t) = H_{I}(t)\Psi_{I}(X_{N+1},t)
		-\mathcal{L}_{N+1}\Psi_I(X_{N+1},t),  
\end{equation}
where $H_I$ is the inner-region Hamiltonian including the Bloch operator.
However, since the Bloch operator only acts on the wavefunction at the boundary, we can
apply the Bloch operator upon the outer-region
wavefunction instead of the inner-region wavefunction:
\begin{equation}
        \label{equation17b}
		i\frac{\partial }{\partial t}\Psi_{I}(X_{N+1},t)=H_{I}(t)\Psi_{I}(X_{N+1},t)
		-\mathcal{L}_{N+1}\Psi(X_{N+1},t).
\end{equation}
We now expand the inner-region wavefunction in terms of the field-free eigenstates
$\psi_k(X_{N+1})$, and
project the inner region TDSE (\ref{equation17}) onto these eigenstates. This projection
then provides a set of equations for the time evolution of the coefficients $C_{k}(t)$:
\begin{equation}
        \label{equation20}
		\left. \frac{d}{dt}C_{k}(t)=-i\sum_{k'}H_{I_{kk'}}(t)C_{k'}(t)
		+\frac{i}{2}\sum_{p}\omega_{pk}\frac{\partial F_{p}(r,t)}{\partial r}\right|_{r=b},
\end{equation}
where $H_{I_{kk'}}(t)$ is the time-dependent Hamiltonian matrix element
between field-free states $\psi_k$ and $\psi_{k'}$, and $\omega_{pk}$ are
the surface amplitudes of eigenstate $\psi_k$ with respect to outer region channel $p$ \cite{Bur11}.

Using the approximation that the time-dependent Hamiltonian is constant within the
time interval $[t,t+\delta t]$, we can now obtain an approximate solution to the
inner-region TDSE (\ref{eq:tdinner}) in terms of the so-called $\phi$
functions \cite{Ska09}. In matrix notation \cite{Ndo09,Lys11,Moo11},
\begin{equation}
        \label{eq:innerprop}
			\mathbf{C}(t+\delta t)\approx e^{-i\delta t\mathbf{H}_{I}}\mathbf{C}(t)
			+\sum _{j=1}^{n_{j}}\left(\delta t\right)^{j}
			\phi _{j}(-i\delta t\mathbf{H}_{I})\mathbf{U}_{j}(t),
\end{equation} 
where
\begin{equation}
        \label{equation25}
		\mathbf{U}_{0}(t)= \mathbf{C}(t), \qquad
		\boldsymbol{U}_{j}(t)=i\frac{d^{j-1}}{dt^{j-1}}\boldsymbol{S}(t),
\end{equation}
with
\begin{equation}
{\bf S}(t) = \frac{1}{2}\sum_{p}\left. \omega_{pk}\frac{\partial F_{p}(r,t)}
{\partial r}\right|_{r=b}.
\end{equation}
The $\phi_j$ functions can be regarded as ``shifted" exponentiation functions, as can be
appreciated through their Taylor series \cite{Lys11},
\begin{equation}
\phi_j(z)= \sum_{k=0}^{\infty} \frac{z^k}{(k+j)!}.
\end{equation}

For the propagation of the inner region wavefunction, we again use an Arnoldi method
\cite{Smy98,Par86}. However, we now need separate propagators to
evaluate each of the $\phi$-functions as well as the $\exp(-i\delta t\mathbf{H}_{I})$
term. Thus each of the ${\bf U}_j$ terms in equation (\ref{eq:innerprop})
is propagated separately, and the final wavefunction at time $t+\delta t$ is obtained
by combining all terms.

\subsection{Description of He and field parameters}

Our interest focuses on the anisotropy parameters in two-colour photoionization
of He and the amount of atomic structure included in the theoretical description. Hence we
describe the He atom using three different basis expansions. These basis expansions can be
characterised by the residual He$^+$-ion states retained in the calculations. The He basis
is built by combining these residual ion states with a complete set of continuum and
bound-state functions for a single outgoing electron. The simplest basis expansion
included employs only the 1s state of He$^+$ as a residual-ion state. Hence the He
basis contains all 1s$n/\varepsilon\ell$ states. The second basis set we employed
including the 1s, 2s and 2p states of He\(^{+}\) as residual-ion states, so that the
He basis contains all 1s$n/\varepsilon\ell$, 2s$n/\varepsilon\ell$ and 2p$n/\varepsilon\ell$
states. For the third basis set, we use pseudo-orbitals, \(\overline{2s}\) and \(\overline{2p}\),
as residual-ion states rather than the physical 2s and 2p orbitals. These pseudo-orbitals are
constructed as Sturmian-type orbitals $r^{i}e^{-\alpha r}$ with the same exponential decay as
the 1s orbital, minimal power of the polynomial term, and orthogonality with respect to the
1s orbital. These basis sets are similar to the ones employed in the investigation of the
choice of gauge for the laser field for time-dependent R-matrix theory \cite{Hut10}.

RMT calculations require a choice for the maximum angular momentum to be included within
the calculations. The calculations using the 1s residual-ion state only were performed
with 3 different values for the maximum angular momentum, $L_{\rm max} = 5$, $L_{\rm max} = 7$,
and $L_{\rm max} = 9$, with the last two calculations producing very similar results.  We
therefore report results for all three basis sets investigated using a maximum angular
momentum $L_{\rm max} = 7$.

In the generation of the He basis, we have used an R-matrix inner region radius of 20 a$_0$.
This inner region is sufficiently large to contain the He ground state. The basis used for
the description of the continuum electron within the inner electron contains 70 B-splines
with order \textit{k} = 11. The knot point distribution varies from a nearly quadratic
spacing near the nucleus to a nearly linear spacing near the inner region boundary. Additional
knot points are inserted further inward to improve the description of functions close
to the nucleus. 
The outer region radial wavefunction for the ejected electron is described on a finite
difference grid extending to 1200 a.u. with a grid spacing
$\delta r$ = 0.075 a$_0$. The time step in the propagation is set to $\delta t$ = 0.005 a.u.
The order of the Arnoldi propagator is 14.

We investigate ionization of helium irradiated by a combination of two laser pulses: an EUV
pulse corresponding to the $17^{\text{th}}$ $-$
$21^{\text{st}}$ harmonic of the fundamental laser field and an overlapping fundamental
dressing field. The wavelength of the fundamental field
ranges from 790~$-$~810 nm and were chosen closely to resemble those used in the
experiment \cite{Hab11}. The IR laser field considered is linearly
polarized in the \textit{z}-direction with an intensity of $5\times10$\(^{10}\) W cm\(^{-2}\)
at peak. The IR pulse profile is given by a 3-cycle $\sin^{2}$
ramp on, followed by 2 cycles at a peak intensity and a 3-cycle $\sin^{2}$ ramp off (3-2-3). 
The EUV laser field considered is linearly polarized in the \textit{z}-direction with a peak
intensity of $1\times10$\(^{11}\) W cm\(^{-2}\). The EUV pulse profile is given by
a 3\textit{n}-cycle  $\sin^{2}$ ramp on, followed by 2\textit{n} cycles at a peak intensity
and a 3\textit{n}-cycle $\sin^{2}$ ramp off (3-2-3), where $n$ indicates the order of the
EUV harmonic. Following the end of the pulse, the wavefunction is propagated in time
corresponding to just over 2.8 cycles of the IR field to ensure that the outgoing electron is
well separated from
the residual ion and to ensure that the ejected electron is indeed a continuum electron.

\section{Results}

In this report, we investigate the asymmetry parameters of two-colour two-photon ionization
of He irradiated by a combination of a short EUV pulse and a short IR pulse using the RMT
approach. These parameters can be obtained from the final-time wavefunction obtained
at the end of the calculations. This wavefunction is described in terms of total angular
momentum, whereas the experimental observations relate primarily to the ejected electron.
In order to compare the results from our investigations to experiment, we therefore first
have to decouple the wavefunction of the outer electron and the wavefunction of the residual
ion states using Clebsch-Gordan coefficients \cite{Har08}. Following this decoupling, we can
construct the  spatial wavefunction for the outgoing electron. This spatial wavefunction is
then transformed into a momentum distribution for the outgoing electron under the assumption
that the Coulomb potential of the residual ion can be neglected.

\begin{figure}[tbh]
	\centering
        \includegraphics[width=8cm]{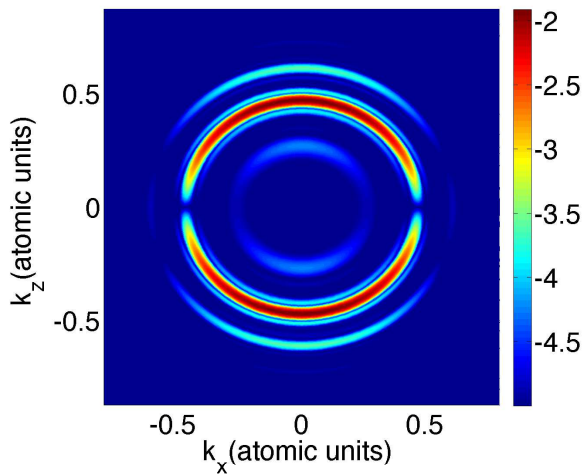}
	\caption{(Colour online) Photoelectron momentum spectrum in the $k_xk_z$ plane for helium irradiated by
	an IR field with wavelength of 790 nm and its 17$^{\text{th}}$ harmonic.}
  \label{fig:momdist}
 \end{figure}

Figure \ref{fig:momdist} shows the ejected-electron momentum distribution of the ejected
electron in the $k_xk_z$-plane for the case that He is irradiated by an IR laser field with
a wavelength of 790 nm and an EUV field given by the 17$^{\rm th}$ harmonic of the IR field.
The IR field is relatively weak, and we therefore show the momentum distribution on a logarithmic
scale. The photoionization spectrum is dominated by the central EUV photoionization peak at
$|k| \approx 0.45$ atomic units. Since the central peak is given by a $p$ outgoing electron,
a node is clearly visible at $k_z=0$ for this central peak. The sidebands generated by absorption
or emission of an additional IR photon can also be identified easily in
figure \ref{fig:momdist}.
%Additional sidebands can be seen corresponding to the absorption
%of two and three IR photons.
%The sideband corresponding to absorption of two IR photons shows again
%a node at $k_z=0$, in contrast to the sidebands corresponding to absorption or emission of a
As will be illustrated later in more detail, these sidebands show no node at $k_z=0$. 

\begin{figure}[tbh]
	\centering
		\includegraphics[width=8cm]{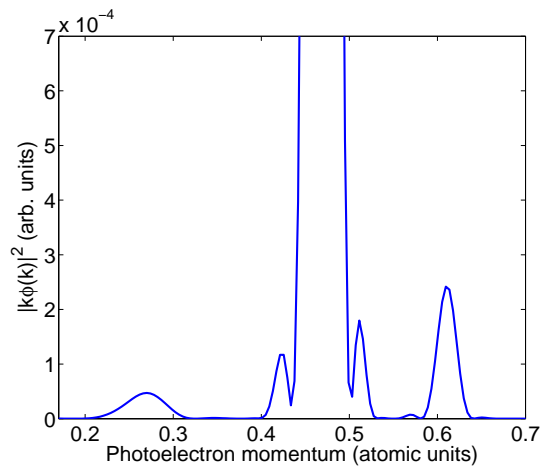} 
	\caption{(Colour online) Photoelectron momentum spectrum for helium irradiated by a 790-nm IR field
	and its 17$^{\text{th}}$ harmonic along the polarization axis of the two laser fields. 
	The momentum spectrum shows that, for an eight-cycle IR-pulse with peak intensity of
	$5\times10^{10}$ W/cm$^2$, the spectrum is dominated by the central EUV peak. Two
	further sidebands can be seen directly by the side of the EUV pulse. These originate
	from the EUV pulse shape. }
  \label{fig:photoe}
 \end{figure}

The relative magnitudes between the central EUV peak and the sidebands can be better assessed 
in figure \ref{fig:photoe}, which shows the photoelectron momentum spectra along the
laser polarization axis for a 790-nm IR field and its 17$^{\rm th}$ harmonic. The figure
shows that for the present laser parameters (short IR pulse, weak intensity), the photoelectron
momentum spectrum is dominated by the central photoemission peak, whereas the sidebands are
significantly weaker in intensity. Two sidebands can be seen immediately by the side of the
main EUV photoelectron peak. These peaks originate from the pulse shape of the EUV laser pulse,
which introduces additional frequency components. The sidebands associated with absorption or
emission of an additional IR photon are well separated from the main EUV peak, and we can thus
easily obtain a total intensity for each of the sidebands at a given emission angle by integration
over the sideband. 

\begin{figure}[tbh]
	\centering
		\includegraphics[width=8cm]{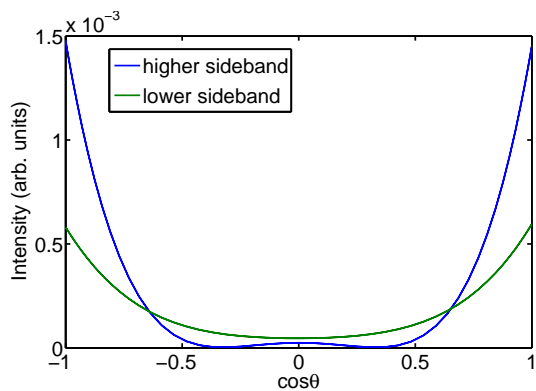} 
	\caption{(Colour online) Intensities of the negative (lower) 
	and positive (higher) sidebands as a function of
	$\cos\theta$ for He irradiated by the 17$^{\text{th}}$ harmonic and overlapping
	fundamental of the 790 nm pulse.}
  \label{fig:angle}
 \end{figure}

The asymmetry parameters for the two-colour two-photon ionization process describe the variation
of the sideband intensity with emission angle, relative to the laser polarization axis, $\theta$.
Figure \ref{fig:angle} shows the variation of the integrated sideband intensity with $\cos\theta$ for
the case of a 790-nm IR field and its 17$^{\rm th}$ harmonic.
The figure demonstrates that the positive sideband (absorption of an IR photon) and the negative
sideband (emission of an IR photon) show significant differences. This difference between positive
and negative sidebands, and hence differences in the asymmetry parameters, has
already been observed before in studies of Ar \cite{Hab09,Hut13}. Figure \ref{fig:angle} shows that
the angular distribution for the higher sideband is more peaked along the laser polarization axis,
whereas the negative sideband shows a more constant behaviour as a function of angle.

This difference between positive and negative sideband is not too unexpected. The present choice
of laser parameters means that the lower sideband corresponds to emission of electrons with an
energy of only 0.8 eV, less than the IR photon energy. At this small energy, interactions with
the residual ion can become important. Furthermore, the centrifugal repulsion potential can
significantly affect the angular distribution. Whereas $s$ electrons do not see this repulsive
potential, $d$ electrons do. The classical turning point for a $d$ electron with an energy
corresponding to the negative sideband is about 2.8 $a_0$, whereas it is about 1.8 $a_0$ at
an energy corresponding to the positive sideband. Hence, the overlap between the $d$ continuum
and the He ground state will be significantly greater for the higher sideband than for the
lower sideband. The emission in the lower sideband should then be more isotropic than the
emission in the higher sideband.

Once we have obtained the angular distributions associated with the sidebands, we can use them
to derive the anisotropy parameters. It is also possible to derive the anisotropy parameters
using a perturbative approach, in which one can derive the anisotropy parameters from the
magnitude and relative phase between the excitation of the $s$ and the $d$ continuum. In the
present time-dependent calculation, we have chosen not to use the latter method, as
$g$ and $\ell=6$ channels are also
populated during the calculations. Since these channels can receive population, they should
be included in the determination of the anisotropy parameters. We also include possible
contributions from $p$, $f$ and $\ell=5$ and 7 channels, even though, for long weak pulses,
they do not contribute to two-photon ionization.

The photoelectron angular distributions for two-photon ionization are given in terms of the
anisotropy parameters by \cite{Hab09}
\begin{equation}
        \label{equation32}
			I(\theta )=\frac{\sigma }{4\pi }[1+\beta _{2}P_{2}(\cos\theta )+
			\beta _{4}P_{4}(\cos\theta )],
\end{equation}
\noindent where $\theta$ is the angle between the laser polarization vector and electron velocity
vector, $\sigma$ is the total cross section, and $\beta _{2}$ and $\beta _{4}$ are the
anisotropy parameters associated with the second and fourth order Legendre polynomials,
respectively. We can thus obtain these parameters by fitting this formula to the theoretical
angular distribution. For the data in figure \ref{fig:angle}, this gives anisotropy parameters
of $\beta_2 =1.83$ and $\beta_4 = 0.48$ for the negative sideband, and $\beta_2= 3.03$ and
$\beta_4= 1.62$ for the positive sideband.

\begin{table*}
\caption{Experimental \cite{Hab11} and theoretical anisotropy parameters for sidebands generated
by the 17$^{\text{th}}$ to the 21$^{\text{st}}$ harmonics of the IR pulse overlapped by the
fundamental pulse in helium. Anisotropy parameters for the lower sidebands are denoted by the
superscript ($-$) whereas positive sidebands are denoted by (+).}
\label{tab:aniso}
\begin{ruledtabular}
\begin{tabular}{c|l|cc|cc} 
Case & Wavelenghts & \(\beta_{2}^{(-)}\) & \(\beta_{4}^{(-)}\) &
\(\beta_{2}^{(+)}\)& \(\beta_{4}^{(+)}\) \\  \hline
 
 A   & 810 nm $17^{\text{th}}$ HH experiment & 2.00$\pm$0.14 &
 0.45$\pm$0.11 & 2.70$\pm$0.14 & 1.12$\pm$0.06 \\ 	% inserting body of the table
 %& 810 nm $17^{\text{th}}$ HH RMT (1 state) & 1.45 & 0.41 & 3.02 & 1.66 \\
    %& 810 nm $17^{\text{th}}$ HH RMT (3 states with real orbitals) & 1.43 & 0.41 & 3.02 & 1.66 \\
    & 810 nm $17^{\text{th}}$ HH RMT (3 states with pseudo-orbitals) &
    1.42 & 0.41 & 3.02 & 1.66 \\ %[1ex]  % [1ex] adds vertical space
\hline 								%inserts single line

B & 801 nm $17^{\text{th}}$ HH experiment & 1.04$\pm$0.09 &
0.30$\pm$0.06 & 2.98$\pm$0.08 & 1.54$\pm$0.06 \\ 		% inserting body of the table
 & 801 nm $17^{\text{th}}$ HH RMT (3 states with pseudo-orbitals) &
 1.65 & 0.45 & 3.02 & 1.64 \\ %[1ex] 				% [1ex] adds vertical space
\hline 								%inserts single line

 C   & 794 nm $17^{\text{th}}$ HH experiment & 1.52$\pm$0.10 &
 0.20$\pm$0.07 & 2.83$\pm$0.08 & 1.38$\pm$0.13 \\ 	% inserting body of the table
  & 794 nm $17^{\text{th}}$ HH RMT (3 states with pseudo-orbitals) &
  1.77 & 0.47 & 3.03 & 1.62 \\ %[1ex]  % [1ex] adds vertical space
\hline 								%inserts single line

  D  & 790 nm $17^{\text{th}}$ HH experiment & 1.73$\pm$0.07 &
  0.55$\pm$0.19 & 2.96$\pm$0.08 & 1.50$\pm$0.03 \\ 			% inserting body of the table
 & 790 nm $17^{\text{th}}$ HH RMT (1 state) & 1.83 & 0.48 & 3.03 & 1.62 \\
  & 790 nm $17^{\text{th}}$ HH RMT (3 states with real orbitals) &
  1.84 & 0.49 & 3.03 & 1.62 \\  
 & 790 nm $17^{\text{th}}$ HH RMT (3 states with pseudo-orbitals) &
 1.84 & 0.49 & 3.03 & 1.62 \\ %[1ex] 				% [1ex] adds vertical space
\hline							%inserts double horizontal lines

 E   & 810 nm $19^{\text{th}}$ HH experiment & 2.07$\pm$0.05 & 0.30$\pm$0.09 &
  2.87$\pm$0.06 & 0.80$\pm$0.03 \\ 					% inserting body of the table
  & 810 nm $19^{\text{th}}$ HH RMT (3 states with pseudo-orbitals) & 2.32 &
  0.67 & 3.03 & 1.49 \\ %[1ex] 				% [1ex] adds vertical space
\hline 							%inserts double horizontal lines

 F  & 790 nm $21^{\text{st}}$ HH experiment & 2.06$\pm$0.05 & 0.13$\pm$0.08 &
 2.43$\pm$0.06 & 0.46$\pm$0.06 \\ 					% inserting body of the table
  & 790 nm $21^{\text{st}}$ HH RMT (3 states with pseudo-orbitals) & 2.55 &
  0.82 & 2.99 & 1.38 \\%[1ex] 				% [1ex] adds vertical space
%\hline 								%inserts single line

\end{tabular}
\end{ruledtabular}
\end{table*}

In the recent experimental investigation of the anisotropy parameters \cite{Hab11}, six combinations
of IR wavelength and associated harmonic were investigated.
For an IR wavelength of 790 nm, and an EUV pulse given by its 17$^{\rm th}$ harmonic,
anisotropy parameters of $\beta_2 = 1.73\pm0.07$ and
$\beta_4= 0.55\pm0.19$ were obtained for the negative sideband, and anisotropy parameters
of $\beta_2 = 2.96\pm0.08$ and $\beta_4 = 1.50\pm0.03$ were obtained for the positive sideband.
Although the theoretical calculations lie slightly outside the experimental error bars for
$\beta_2$ for the negative sideband and $\beta_4$ for the positive sideband, the overall
agreement for this choice of laser wavelengths is very good. A more detailed comparison of
the theoretical and experimental anisotropy parameters is given in table \ref{tab:aniso}, where
anisotropy parameters are presented for all combinations reported experimentally.  

For the case discussed above, an IR wavelength of 790 nm, and an EUV pulse given by its
17$^{\rm th}$ harmonic, we also show in table \ref{tab:aniso}, a comparison of the
anisotropy parameters obtained using the different He basis sets. The table shows that
the differences in the anisotropy parameters with respect to basis size are minimal in
the present calculations, with a change of 0.01 in $\beta_2$ for the negative sideband.
This level of change is typical: the largest difference is seen for case A, where $\beta_2$
for the negative sideband increases to 1.45 when only the 1s state of He$^+$ is included.
The changes between the different basis sets lie thus within the experimental uncertainty.
We have therefore chosen to present results only for the case where the He basis is constructed
using the He$^+$ 1s, $\overline{\rm 2s}$ and $\overline{\rm 2p}$ residual-ion states.

The anisotropy parameters calculated using the RMT method, shown in table \ref{tab:aniso},
are noticeably smaller in magnitude for the negative
sideband than for the positive sideband. $\beta_{2}^{(+)}$ is very similar in all cases studied,
with a value very close to 3.  $\beta_{2}^{(-)}$ shows greater variation: it increases with
increasing photon energy of the harmonic. The negative sideband straddles the ionization
threshold at a wavelength of 810 nm. Just above the ionization threshold, ionization will be
dominated by $s$ electrons, and therefore the closer the negative sideband gets to the
ionization threshold, the emission process should become more and more isotropic. However, it
may be difficult to obtain accurate anisotropy parameters very close to threshold in the present
calculations, as the wavefunction may need to be propagated for long times to separate population
in high Rydberg states from population in the low-energy continuum. The $\beta_{4}$ parameters
show variation in both cases: we find a relatively small decrease in $\beta_{4}^{(+)}$  with
increasing EUV energy, whereas the $\beta_{4}^{(-)}$ shows a steady increase.  %\\

We can compare our anisotropy parameters with predictions using the Soft Photon Approximation
(SPA) \cite{Maq07}. In this approximation, it is assumed that the absorption of the IR photon
does not significantly affect the energy of the outgoing electron. This approximation may not
work
well for the 17$^{\rm th}$ harmonic due to its proximity to the ionization threshold, but the
approximation should be more appropriate for higher
harmonics. If it is assumed that the IR intensity is weak, then simple analytic expressions can
be derived for the anisotropy parameters, in terms of
the single-photon anisotropy parameter, estimated at the energy of the sideband \cite{Hab11}:
\begin{equation}
	\beta_{2}^{\pm }=\frac{5}{7}\left ( \frac{14+11\beta_{2}^{(0)}}{5+2\beta_{2}^{(0)}}
	\right ), \beta_{4}^{\pm }=\frac{36}{7}\left ( \frac{\beta_{2}^{(0)}}{5+2\beta_{2}^{(0)}}
	\right ).
\label{eq:spa}
\end{equation}
For single-photon ionization of the He ground-state to just above the He$^+$ 1s threshold,
only $p$ electrons can be ejected, and therefore the
parameter $\beta_2^{(0)} = 2$ at all photon energies considered here. Substitution of this
value into the SPA estimates for the anisotropy
parameters gives the following values, $\beta_{2}^{\pm }~=~20/7\approx2.86$ and
$\beta_{4}^{\pm }~=~8/7\approx1.14$. The values
shown in table \ref{tab:aniso} show that with increasing EUV-photon energy the anisotropy
parameters change towards the SPA predictions,
but that higher EUV-photon energies than studied here need to be considered for the SPA to
provide close agreement. 

Table \ref{tab:aniso} also provides a comparison of the experimentally obtained anisotropy
parameters with the present ones. The agreement
between theory and experiment is best for the case studied earlier: an IR field of 790 nm
and the 17$^{\rm th}$ harmonic, case D in table \ref{tab:aniso}. For the other cases,
significant differences are seen between the experimental results and the present theoretical
results. The best agreement is seen for $\beta_2^{(+)}$, but for $\beta_2^{(-)}$, other than
case D, the smallest difference is 0.25. For the $\beta_4$ parameters, reasonably good
agreement is obtained for all cases involving the 17$^{\rm th}$ harmonic, but the differences
become pronounced for the cases involving the 19$^{\rm th}$ and
21$^{\rm st}$ harmonic. 

The origin of the differences between the experimental results and theory is unclear.
The $\beta_4$ parameters can be related directly to
the relative magnitude between the emission of a $d$ electron and emission of an $s$
electron \cite{Hab11}. The present results, in particular those
for the negative sideband, are consistent with a picture in which the emission of $d$
electron is reduced when one approaches the ionization
threshold. Extrapolation of the numerical results to higher photon energy gives
anisotropy parameters consistent with the predictions of the
SPA. On the other hand, the present calculations use very clean laser pulses for
both the IR and the EUV pulse. It is unrealistic to expect
such a clean pulse for the EUV pulse experimentally, as it is obtained through
harmonic generation. These differences in laser parameters may
well be the root origin for the differences seen between theory and experiment. 

\section{Conclusions}

We have demonstrated the application of the RMT approach to the investigation of photoelectron
angular distributions and anisotropy
parameters derived from these angular distributions. The negative sidebands tend to be more
isotropic than the positive sidebands. The obtained anisotropy parameters differ noticeably
from the predictions of the SPA due to the proximity of the EUV photon energy to the ionization
threshold. For increasing EUV photon energy, the anisotropy parameters move closer to
the predictions of the SPA. The agreement with experiment is good for the case of an IR pulse
with a wavelength of 790 nm and an EUV pulse given by its 17$^{\rm th}$ harmonic, but notable
differences are seen for other combinations of laser pulse. 

The RMT approach has been developed recently for the investigation of atomic processes in
intense ultra-short light fields. The present
calculations demonstrate that the approach can be used to obtain photoelectron distributions.
In the present study, we compare these
distributions for the case that the IR field is weak, but the approach should also be capable
of treating more intense IR fields. In these cases,
the sideband structure becomes significantly more complicated \cite{Mey13}, and it would be
interesting to see how the approach compares to, for
example, the SPA when more IR photons are absorbed by the ejected electron. Whereas calculations
at low IR intensity can be carried out using either
the TDRM or RMT approach, at high IR intensities, the RMT approach would be strongly preferred, as
the RMT approach is more suitable for large-scale parallelisation. This increase in parallelisation
scale becomes particularly important when many angular momenta need to be included in the
calculations.

\section*{Acknowledgements}

This research has been supported by the European Commission
Marie Curie Initial Training Network CORINF and by
the UK Engineering and Physical Sciences Research Council under grant no. EP/G055416/1. The
authors would like to thank J.S. Parker and K.T. Taylor for assistance with the RMT codes and
valuable discussions. The main development of the RMT codes was carried out by M.A. Lysaght
and L.R. Moore. This work made use of the facilities of HECToR, the UK's national high-performance computing service,
which was provided by UoE HPCx Ltd at the University of Edinburgh, Cray Inc and NAG Ltd, and
funded by
the Office of Science and Technology through EPSRC's High End Computing Programme.


\begin{thebibliography}{99}


\bibitem{Iva09}
M. Ivanov and F. Krausz, {\em Rev. Mod. Phys. } {{\bf 81}, 1634 (2009)}.
\bibitem{Pau01}
P.M. Paul, E.S. Toma, P. Breger, G. Mullot, F. Aug\'{e},
P. Balcou, H.G. Muller and P. Agostini, 2001 {\it Science\/} {\bf 292} 1689
(2001)
\bibitem{Ven96}
V. Veniard, R. Ta\"\i eb and A. Maquet, {\em Phys. Rev. A} {\bf 54}, 721 (1996)
%\bibitem{Ago04}
%P. Agostini and L. F. DiMauro, {\em Rep. Prog. Phys.} {{\bf 67}, 813 (2004)}.
\bibitem{A13}
M. Yabashi \textit{et al}, \textit{J. Phys. B} {\bf 46}, 164001 (2013)
\bibitem{B13}
J. Feldhaus \textit{et al}, \textit{J. Phys. B} {\bf 46}, 164002 (2013)
\bibitem{C13}
C. Bostedt \textit{et al}, \textit{J. Phys.B} {\bf 46}, 164003 (2013).
\bibitem{Mey13}
S. D\"usterer, L. Rading, P. Johnsson, A. Rouzee, A. Hundertmark, M.J.J. Vrakking,
P. Radcliffe, M. Meyer, A.K. Kazansky and N.M. Kabachnik, {\em J. Phys. B} {\bf 46}, 164026 (2013)
\bibitem{Maq07}
A. Maquet and R. Ta\"{i}eb, {\em J. Mod. Opt.} {{\bf 54}, 1847 (2007)}.
\bibitem{Nik08}
L.A.A. Nikolopoulos, J.S. Parker and K.T. Taylor, {\em Phys. Rev. A} {\bf 78}, 063420 (2008).
\bibitem{Moo11}
L. R. Moore, M.A. Lysaght, L.A.A. Nikolopoulos, J.S. Parker, H.W. van der
Hart and K.T. Taylor, {\em J. Mod. Opt.} {{\bf 58}, 1132 (2011)}.
\bibitem{Lys11}
M. A. Lysaght, L. R. Moore, L. A. A. Nikolopoulos, J. S. Parker, H. W. van der Hart, and
K. T. Taylor {\em Quantum Dynamic Imaging: Theoretical and Numerical Methods} Eds. A. D.
Bandrauk and M. Ivanov (Springer, Berlin, 2011), pp. 107-134.
\bibitem{Mey10}
M. Meyer, J.T. Costello, S. D\"usterer, W.B. Li and P. Radcliffe,
{\em J. Phys. B} {\bf 43}, 194006 (2010).
\bibitem{Hab11}
L. H. Haber, B. Doughty, and S. R. Leone, {\em Phys. Rev. A} {{\bf 84}, 013416 (2011)}.
\bibitem{Mon13}
S. Mondal \textit{et al}, {\em J. Phys. B: At. Mol. Opt. Phys.} {\bf 46}, 205601 (2013)
\bibitem{OKe13}
P. O'Keefe, A. Mihelic, P. Bolognesi, M. zitnik, A. Moise, R. Richter and L. Avaldi, 
{\em New. J. Phys.} {\bf 15}, 013023 (2013)
\bibitem{Hut13}
S. Hutchinson, M. A. Lysaght and H. W. van der Hart, {\em Phys. Rev. A} {{\bf 88}, 023424 (2013)}.
\bibitem{Hab09} 
L. H. Haber, B. Doughty, and S. R. Leone, {\em J. Phys. Chem.} {{\bf A 113}, 13152 (2009)}.
\bibitem{Tom01}
E.S. Toma and H.G. Muller, {\em J. Phys. B} {\bf 35}, 3435 (2002).
\bibitem{Moo11b}
L. R. Moore, M.A. Lysaght, J.S. Parker, H.W. van der Hart and K.T. Taylor,
{\em Phys. Rev. A} {{\bf 84}, 061404 (2011)}.
\bibitem{Har14}
H.W. van der Hart, {\em Phys. Rev. A} {\bf 89}, 053407 (2014).
\bibitem{Har14b}
H.W. van der Hart and R. Morgan, {\em Phys. Rev. A} {\bf 90}, 013424 (2014).
\bibitem{Lys09}
M. A. Lysaght, H.W. van der Hart and P.G. Burke, {\em Phys. Rev. A} {{\bf 79}, 053411 (2009)}.
%\bibitem{Lys09b}
%M. A. Lysaght et al, {\em Phys. Rev. Lett.} {{\bf 102}, 193001 (2009)}.
\bibitem{Smy98}
E.S. Smyth, J.S. Parker and K.T. Taylor, {\em Comp. Phys. Comm.} {{\bf 114}, 1 (1998)}.
\bibitem{Par86}
T. J. Park and J. C.Light,  {\em J. Chem. Phys.} {{\bf 85}, 5870 (1986)}.
%\bibitem{Wig46}
%E. P. Wigner. {\em Phys. Rev.}, {{\bf 70}, 15-33 (1946)}.
%\bibitem{Wig47} 
%E. P. Wigner and L. Eisenbud. {\em Phys. Rev.},{{\bf 72}, 29-41 (1947)}.
\bibitem{Bur11} 
P. G. Burke in {\em R-Matrix in Theory of Atomic Collisions}, Springer Series on Atomic,
Optical, and Plasma Physics, Vol. 61 (2011).
\bibitem{Ska09}
B. Skaflestad and W.M. Wright, {\em Appl. Numer. Math.} {\bf 58}, 783 (2009)
\bibitem{Ndo09}
M. Ndong, H. Tal-Ezer and R.K.C.P. Koch, {\em J. Phys. Chem.} {\bf 130}, 124108 (2009).
\bibitem{Hut10} 
S. Hutchinson, M. A. Lysaght and H. W. van der Hart, {\em J. Phys.} {{\bf B 43}, 095603 (2010)}.
\bibitem{Har08} 
H.W. van der Hart, M. A. Lysaght, and P. G. Burke, {\em Phys. Rev.} {{\bf A 77}, 065401 (2008)}.
%\bibitem{Rei03} 
%K. L. Reid. {\em Annu. Rev. Phys. Chem.}, {{\bf 54}, 397-424 (2003)}.
\end{thebibliography}
\end{document}